\documentclass[
    manuscript,
    nonacm, % disables ACM’s copyright/DOI blocks
    acmsmall,
    natbib=false,
    screen,
]{acmart}
\AtBeginDocument{
  }

\RequirePackage[
    datamodel=acmdatamodel,
    style=acmnumeric,
]{biblatex}
\usepackage{listings}
\lstdefinelanguage{Julia}
  {morekeywords={abstract,break,case,catch,const,continue,do,else,elseif,
      end,export,false,for,function,immutable,import,importall,if,in,
      macro,module,otherwise,quote,return,switch,true,try,type,typealias,
      using,while},
   sensitive=true,
   alsoother={$},
   morecomment=[l]\#,
   morecomment=[n]{\#=}{=\#},
   morestring=[s]{"}{"},
   morestring=[m]{'}{'},
}[keywords,comments,strings]

\usepackage{subcaption}

\begin{document}

\title{Efficient Symbolic Computation via Hash Consing}

\thanks{Under review.}

\author{Bowen Zhu}
\email{bowenzhu@mit.edu}
\affiliation{
    \institution{MIT}
    \department{EECS}
    \department{CSAIL}
    \city{Cambridge}
    \state{MA}
    \country{USA}
}

\author{Aayush Sabharwal}
\email{aayush.sabharwal@juliahub.com}
\affiliation{
    \institution{JuliaHub}
    \country{India}
}

\author{Songchen Tan}
\email{songchen@mit.edu}
\affiliation{
    \institution{MIT}
    \department{Math}
    \department{CSAIL}
    \city{Cambridge}
    \state{MA}
    \country{USA}
}

\author{Yingbo Ma}
\email{mayingbo5@gmail.com}
\affiliation{
    \institution{Independent Researcher}
    \country{USA}
}

\author{Alan Edelman}
\email{edelman@mit.edu}
\affiliation{
    \institution{MIT}
    \department{Math}
    \department{CSAIL}
    \city{Cambridge}
    \state{MA}
    \country{USA}
}

\author{Christopher Rackauckas}
\email{crackauc@mit.edu}
\affiliation{
    \institution{MIT}
    \department{CSAIL}
    \city{Cambridge}
    \state{MA}
    \country{USA}
}

\renewcommand{\shortauthors}{Bowen Zhu, Aayush Sabharwal, Songchen Tan, Yingbo Ma, Alan Edelman, Chris Rackauckas}

\begin{abstract}
Symbolic computation systems suffer from memory inefficiencies due to redundant storage of structurally identical subexpressions, commonly known as \emph{expression swell}, which degrades performance in both classical computer algebra and emerging AI-driven mathematical reasoning tools. In this paper, we present the first integration of hash consing into JuliaSymbolics, a high-performance symbolic toolkit in Julia, by employing a global weak-reference hash table that canonicalizes expressions and eliminates duplication. This approach reduces memory consumption and accelerates key operations such as differentiation, simplification, and code generation, while seamlessly integrating with Julia's metaprogramming and just-in-time compilation infrastructure. 
Benchmark evaluations across different computational domains reveal substantial improvements: symbolic computations are accelerated by up to 3.2×, memory usage is reduced by up to 2×, code generation is up to 5× faster, function compilation up to 10× faster, and numerical evaluation up to 100× faster for larger models. While certain workloads with fewer duplicate unknown-variable expressions show more modest gains or even slight overhead in initial computation stages, downstream processing consistently benefits significantly.    
These findings underscore the importance of hash consing in scaling symbolic computation and pave the way for future work integrating hash consing with e-graphs for enhanced equivalence-aware expression sharing in AI-driven pipelines.
\end{abstract}

\maketitle

\section{Introduction}
Symbolic computation systems serve as fundamental tools across various domains including computer algebra, program analysis, automated theorem proving, scientific computing, and AI-driven mathematical reasoning \cite{GröbnerBases1998,EfficientAlgoGrobner,ToRA2024}. These systems manipulate complex mathematical expressions represented as abstract syntax trees (ASTs) to perform algebraic transformations, simplifications, and evaluations \cite{ReviewMathematica1992}. 

However, as symbolic workloads grow in scale—from classical tasks like symbolic differentiation or polynomial manipulation to modern challenges in tool-integrated AI systems—the redundancy of structurally identical subexpressions becomes a critical bottleneck. This phenomenon, known as \textit{expression swell} \cite{AlgoComputerAlgebra}, leads to increased memory consumption, frequent garbage collection (GC) cycles, and degraded cache performance, ultimately affecting computational efficiency.

Hash consing, a technique first introduced by \textcite{HashConsing1958} and adopted in Lisp systems \cite{MonocopyLisp}, addresses this problem by ensuring structurally equivalent expressions share a single memory allocation. By canonicalizing terms through hash-based interning, hash consing eliminates redundancy while enabling constant-time structural equality checks—a method later adopted in theorem provers (e.g. ACL2 \cite{ACL2}) and compiler intermediate representations \cite{IntermediateLang}. Recent advances in AI for mathematics, such as ToRA \cite{ToRA2024} and MuMath-Code \cite{MuMath-Code2024}, have further highlighted the need for efficient symbolic primitives: these systems combine large language models (LLMs) with symbolic backends, requiring rapid interleaving of neural-guided code generation and term manipulation.

While effective in statically-typed ecosystems, integrating hash consing into dynamically-typed symbolic systems (e.g. Julia, Python) poses unique challenges, including balancing immutability with dynamic dispatch, minimizing GC overhead \cite{hash-consing-GC}, and meeting the latency requirements of modern neuro-symbolic pipelines. The success of systems like NuminaMath \cite{NuminaMath2024}, which won the 2024 AIMO Progress Prize \cite{AIMO2024} by solving competition-level problems through iterative LLM-symbolic collaboration, demonstrates how optimizing symbolic infrastructure directly impacts hybrid AI systems' ability to tackle advanced mathematics.

In this paper, we present the first integration of hash consing into JuliaSymbolics \cite{JuliaSymbolics}, a high-performance symbolic toolkit in the Julia language \cite{Julia2017}. Experimental results show that for the B Cell Receptor (BCR) signaling model—comprising 1,122 ODEs and 24,388 reactions—our approach improves symbolic Jacobian computation time by approximately 3.2×, reduces memory requirements by around 2×, and accelerates both function compilation and evaluation by about 2×. In contrast, for a thermal fluid dynamics model based on the X Steam thermodynamic system, the Jacobian computation is slightly slower (due to fewer duplicated expressions in unknown variables), and memory usage increases by about 2.5×. However, code generation is consistently improved by approximately 5×, function compilation is nearly 10× faster, and function evaluation is accelerated by 20× to 100× depending on model size.

\textbf{Contributions:}
\begin{enumerate}
    \item A hash consing framework for JuliaSymbolics that integrates transparently into existing symbolic workflows.
    \item Comprehensive benchmarks on ODE modeling and simulation (\autoref{sec:evaluation}), quantifying significant reductions in memory usage (e.g., $2\times$ for BCR Jacobian computation) and substantial runtime speedups across various stages, including symbolic computation ($3\times$ for BCR), compilation (up to $10\times$ for XSteam), and numerical evaluation (up to $100\times$ for XSteam), particularly for large-scale problems.
    \item Design principles and lessons learned for adopting hash consing in symbolic computation systems, offering insights applicable to other languages and toolchains.
\end{enumerate}

This work bridges a critical gap between classical optimization techniques and next-generation symbolic toolchains. Our results underscore the relevance of hash consing in scaling symbolic methods to large-scale problems in scientific computing, formal methods, and neuro-symbolic AI, while providing actionable insights for improving symbolic infrastructure in dynamically-typed environments.

\section{Background and Motivation}
Symbolic computation systems manipulate mathematical expressions in their symbolic form rather than operating on numerical values \cite{ComputerAlgebra2003}. These systems represent mathematical objects, including variables, functions, operators, and constants, as structured data, typically organized in expression trees.
In most systems, these expression trees are implemented as object graphs, where each node is a distinct object with pointers to its children. This approach offers flexibility in manipulating expressions but introduces significant memory overhead, particularly for large expressions. 

As symbolic manipulations are performed—whether in classical computer algebra or modern AI-driven workflows—identical subexpressions are frequently regenerated. Expression transformation algorithms often create multiple variants of an expression before selecting the optimal form. During this process, identical subexpressions are constructed repeatedly. \textcite{ComparingSpeedPolynomial2003} demonstrated that computer algebra systems may regenerate the same subexpressions dozens of times during complex simplification procedures. The same occurs in symbolic differentiation. For example, computing the derivative of 
\begin{equation}
    f(x)=64x(1-x)(1-2x)^2\left(1-8x+8x^2\right)^2
\end{equation}
yields
\begin{equation}
    \begin{split}
        f'(x)=&128x(1-x)(-8+16x)(1-2x)^2\left(1-8x+8x^2\right)
        +64(1-x)(1-2x)^2\left(1-8x+8x^2\right)^2\\
        &-64x(1-2x)^2\left(1-8x+8x^2\right)^2
        -256x(1-x)(1-2x)\left(1-8x+8x^2\right)^2
    \end{split}
\end{equation}
which contains many duplicate subterms.

In emerging AI-driven workflows, such as tool-integrated reasoning agents \cite{ToRA2024,MuMath-Code2024}, these redundancies are exacerbated. Systems like ToRA and MuMath-Code iteratively generate symbolic code snippets via large language models (LLMs), verify their correctness with symbolic backends, and refine expressions through repeated manipulation. The state-of-the-art NuminaMath system~\cite{NuminaMath2024}, which won the 2024 AIMO Progress Prize by solving 29/50 competition-level problems, exemplifies how LLM-symbolic collaboration requires frequent construction, comparison, and elimination of intermediate terms—processes fundamentally impacted by expression swell.

Memory redundancy impacts performance in several critical ways across both classical and AI-driven workflows:
\begin{itemize}
    \item \emph{Increased memory consumption}: The most direct effect is excessive memory usage, which can limit the scale of problems that can be addressed.
    \item \emph{Cache inefficiency}: Redundant expressions spread across memory reduce cache locality, degrading performance on modern CPU architectures.
    \item \emph{Garbage collection overhead}: As redundant expressions are created and discarded, garbage collection cycles become more frequent and time-consuming.
    \item \emph{Repeated computation}: Without expression sharing, identical subexpressions must be reevaluated repeatedly.
    \item \emph{Expensive equality testing}: Structural comparison of expressions becomes unnecessarily expensive when physical equality could suffice.
    \item \emph{Inefficient code generation}: Explicit common subexpression elimination (CSE) has to be applied to avoid redundant computation in code generation.
\end{itemize}

Hash consing is a technique that directly targets the problem of redundancy by ensuring that every distinct immutable structure is stored exactly once in memory. The core idea is to maintain a global table of previously allocated structures. When a new structure is constructed, the system checks the table for an equivalent structure. If one is found, the existing instance is reused; otherwise, the new structure is added to the table. This method, known as maximal sharing, allows equality checks to be performed by simple pointer comparison rather than a full structural traversal, greatly enhancing efficiency. \textcite{ImplementingFastEquality1994} demonstrated that hash consing could be systematically applied in functional programming to accelerate equality checks and optimize data structures such as sets and maps, tools that are central to many symbolic systems. 

Modern AI-driven symbolic workflows inherit these benefits while introducing new requirements. For example, systems like MuMath-Code \cite{MuMath-Code2024} generate hundreds of candidate expressions via LLMs, requiring rapid equivalence checks against ground-truth solutions. Hash consing enables constant-time equality verification during this filtering process. Similarly, in theorem provers like ACL2 \cite{ACL2} or Rocq \cite{HashConsCoq2014}, hash consing has long been used to manage proof term complexity—a capability now critical for verifying LLM-generated derivations in frameworks like ToRA \cite{ToRA2024}. The adoption of hash consing in compilers (e.g., LLVM’s global value numbering \cite{LLVM:CGO04}) further demonstrates its versatility. These motivate our work to adapt hash consing for modern symbolic toolchains, ensuring its benefits extend to both classical computer algebra and emerging AI-driven mathematical reasoning.

\section{Hash Consing in JuliaSymbolics}\label{sec:hash-consing}
Most traditional symbolic computation systems, such as Mathematica and Maple, are designed primarily for general-purpose algebraic manipulation, theorem proving, and exact symbolic simplification \cite{Mathematica,Maple}. Although these systems offer extensive symbolic capabilities, they are not suitable for the specific demands of modern scientific computing, where symbolic expressions must be efficiently transformed into high-performance numerical code.

JuliaSymbolics \cite{JuliaSymbolics} distinguishes itself by focusing on symbolic-numeric integration, particularly in the context of differential-algebraic equations (DAEs). Unlike conventional symbolic tools that treat expressions as static mathematical objects, JuliaSymbolics emphasizes efficient symbolic transformations that enhance numerical performance. This is particularly useful in applications such as scientific modeling, where large-scale symbolic systems need to be compiled into highly optimized numerical routines.

Another advantage of JuliaSymbolics is its role in facilitating model transformations and code generation \cite{ModelingToolkit2021}. By leveraging Julia’s just-in-time (JIT) compilation and multiple dispatch \cite{Julia2017}, JuliaSymbolics enables seamless conversion of symbolic expressions into efficient runtime-executable code. This makes it particularly suitable for applications where symbolic expressions must be dynamically modified and evaluated within a high-performance computing environment.

However, the original implementation of JuliaSymbolics suffered from two main issues:
\begin{itemize}
    \item \emph{Memory overhead}: The representation of symbolic expressions in JuliaSymbolics was not fully optimized. Redundant storage of subexpressions leads to significant memory consumption. This is particularly problematic when dealing with complex differential-algebraic systems, where even minor redundancies can accumulate and hinder scalability.
    \item \emph{Inefficient code generation}: While one of the primary goals of JuliaSymbolics is to facilitate model transformations and generate optimized numerical code, the code generation pipeline had room for improvement, particularly on common subexpression elimination (CSE). The inefficiencies in this process translated to slower runtime performance and can negate the benefits of symbolic preprocessing in high-performance applications.
\end{itemize}

To address memory redundancy in symbolic workflows, we integrated hash consing into JuliaSymbolics by modifying the core term-construction logic in SymbolicUtils.jl, the foundational library for symbolic manipulation. Our implementation ensures structural sharing of symbolic objects while maintaining compatibility with existing simplification, differentiation, and code generation pipelines.

\subsection{Utilizing Weak-Reference Dictionary for Memory Management}
A global
% thread-safe 
hash table is central to the hash consing mechanism. To balance memory efficiency with lookup performance, we employed weak references for storing symbolic objects. This design allows the garbage collector (GC) to reclaim memory for terms that are no longer strongly referenced elsewhere in the program, preventing unbounded growth of the hash table.

For this purpose, we used the \texttt{WeakValueDict}, a Julia-native weak-reference dictionary \cite{WeakValueDicts}. This data structure ensures that once an object is no longer in use elsewhere in the program, it can be garbage collected, even if it remains in the dictionary. This strategy maintains the balance between efficient object retrieval and optimal memory management. 

Each term construction requires a hash computation and table lookup. While hash-based retrieval is amortized constant time, recursive hashing of deep expression trees adds non-trivial latency. So we cache hash for symbolic terms to reduce redundant calculation in deep traversals.

Hash collisions pose a correctness risk: two distinct expressions may hash to the same value, leading to incorrect sharing. To mitigate this, we fallback to structural checks on hash collisions.

\subsection{Modifying Constructors in SymbolicUtils.jl}
The core functionality of JuliaSymbolics is built upon the SymbolicUtils.jl package. To implement hash consing seamlessly, we modified the constructors of \texttt{BasicSymbolic} objects within this package. Upon creation of a new symbolic object, the constructor computes its hash and checks the global hash table for an existing equivalent object. If a match is found, the existing object is returned; otherwise, the new object is added to the table and returned. This modification ensures that hash consing is applied transparently across all operations in SymbolicUtils.jl and its downstream packages. The process is transparent to users, requiring no changes to existing code that constructs or manipulates symbolic expressions.

\subsection{Simplifying Directed Acyclic Graphs that Represent Expressions}
\begin{figure}[hbt!]
    \centering
    \begin{subfigure}{0.59\textwidth}
        \centering
        \includegraphics[height=5cm]{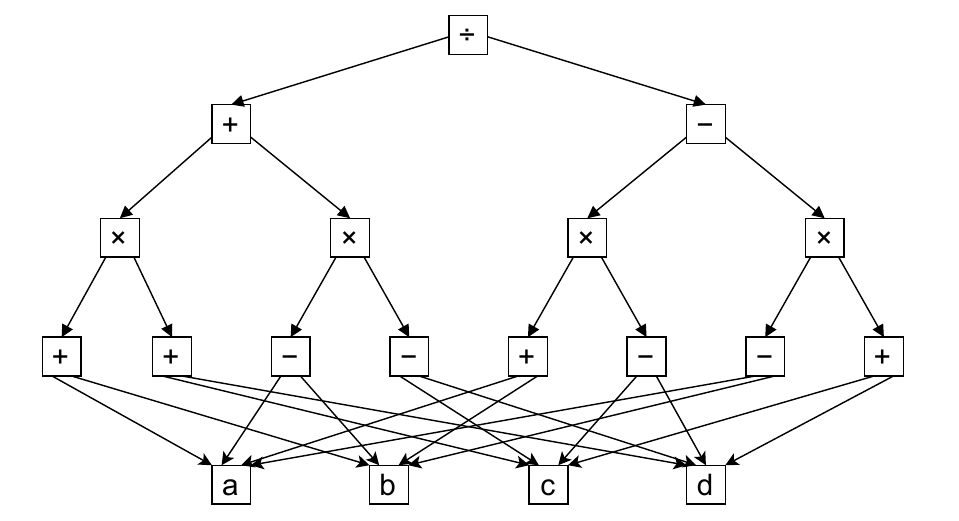}
        \Description{A graph showing the structure without hash consing}
        \caption{Representation without hash consing}
        \label{fig:tree}
    \end{subfigure}
    \begin{subfigure}{0.4\textwidth} 
        \centering
        \includegraphics[height=5cm]{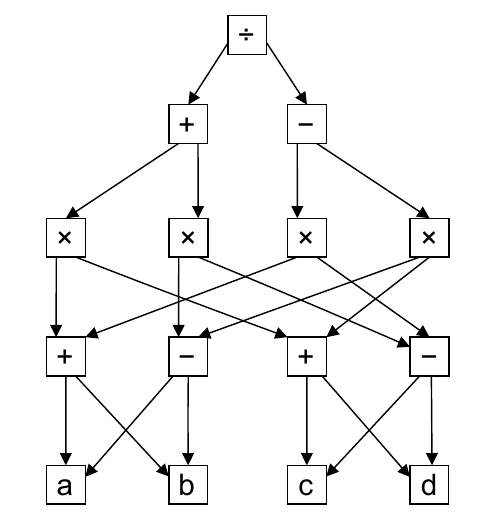}
        \Description{A graph showing the structure with hash consing}
        \caption{Representation with hash consing}
        \label{fig:DAG}
    \end{subfigure}
    \caption{Two types of representations for $\displaystyle\frac{(a + b)(c + d) + (a - b)(c - d)}{(a + b)(c - d) - (a - b)(c + d)}$}
    \label{fig:two_images}
\end{figure}

By implementing hash consing, the directed acyclic graphs (DAGs) that represents symbolic expressions in JuliaSymbolics could be simplified. For instance, the expression constructed with the following Julia code
\begin{lstlisting}
using Symbolics
@variables a b c d
x = (a + b) * (c + d)
y = (a - b) * (c - d)
z = (a + b) * (c - d)
w = (a - b) * (c + d)
term = (x + y) / (z - w)
\end{lstlisting}
has duplicate terms in its graph as shown in \autoref{fig:tree}, since the same expression is constructed multiple times throughout the computation. With hash consing, during the construction of an expression, it is checked against the global weak-reference dictionary; if the same expression already exists, the system will use a pointer to the existing expression instead of creating a duplicated one, leading to more efficient memory usage and faster computations, as shown in \autoref{fig:DAG}. 

\subsection{Optimizing Code Generation}
Hash consing enables a more efficient code generation pipeline by transforming the internal symbolic representation from a tree to a directed acyclic graph (DAG), where structurally identical subexpressions are unified into a single node. In traditional symbolic systems, code generation must explicitly perform common subexpression elimination (CSE) to identify and factor out redundancy. However, in our implementation, this redundancy is inherently removed by hash consing at construction time. To take full advantage of this new DAG structure, we modified the existing code generation algorithm in JuliaSymbolics, which was originally designed for tree-based representations and lacked awareness of shared substructures. Our revised algorithm performs a simple topological sort over the DAG to determine evaluation order, automatically avoiding redundant computations and generating more compact and efficient numerical code. This not only reduces code generation time for large expressions but also simplifies the overall pipeline by eliminating the need for separate CSE passes.

\subsection{Implementing Memoization}
To further optimize performance, we implemented memoization for specific functions within JuliaSymbolics. Memoization stores the results of expensive function calls within each individual thread so that when the same inputs occur again, the cached result is returned instead of recomputing the function, thus reducing redundant computations and improving execution speed. Importantly, our hash consing framework guarantees that structurally equivalent expressions are represented by the same object in memory. This property enables memoization to be applied reliably, as it ensures that identical inputs are always recognized as such. 

\section{Evaluation}\label{sec:evaluation}
This section presents a benchmark comparison of computational performance for two distinct problems: a B cell receptor (BCR) signaling model and a thermal fluid dynamics problem using the X Steam library. The evaluation focuses on the time and memory requirements for Jacobian symbolic computation, the time for code generation, numerical function compilation and execution across varying model sizes.

\subsection{Problem Descriptions}
\subsubsection{B Cell Receptor (BCR) Signaling Model}\label{sec:BCR}
The BCR signaling model \cite{BCRModel} comprises a system of ordinary differential equations (ODEs) derived from a rule-based description of BCR signaling. It captures the membrane-proximal interactions of six proteins (BCR, Lyn, Fyn, Csk, PAG1, and Syk), resulting in 1,122 ODEs and 24,388 reactions. Each ODE follows the form:
\begin{equation}
    \frac{\mathrm d[\text{Species}]}{\mathrm dt}=\sum\left(\text{Production terms}\right)-\sum\left(\text{Decay terms}\right)
\end{equation}
where production terms are binding, phosphorylation, or complex formation and decay terms are dissociation, dephosphorylation, or degradation. The BCR model uses Michaelis-Menten and Hill function style representations of switches in order to represent nonlinear feedback mechanisms and regulations. This model is computationally intensive and relevant to JuliaSymbolics' focus on differential equations and code generation for systems biology.

\subsubsection{X Steam Thermodynamic System}\label{sec:thermal}
The XSteam.jl library \cite{XSteam.jl}, a Julia implementation of the IAPWS IF97 standard formulation for the thermodynamic properties of water and steam \cite{IAPWS-IF97}, provides a suite of functions for calculating various thermodynamic properties, such as density, enthalpy, entropy, and heat capacity, given pressure and temperature. While the original XSteam functions are written for numerical evaluation, we constructed a symbolic representation of a simplified thermodynamic system using a selection of XSteam functions. We systematically varied the complexity of this thermodynamic system by increasing the number of state variables and thermodynamic properties included in the model. This allowed us to analyze how hash consing scales with increasing model size. The symbolic expressions generated in this process are large and complex, providing a challenging test case for expression manipulation and simplification.

\subsection{Experimental Setup}
The benchmarks were performed on a system with the following specifications: Ubuntu Linux (GNU/Linux), x86\_64 architecture, Kernel version 5.15.0-58-generic operating system and AMD EPYC 7502 32-Core Processor. The system had sufficient RAM to ensure that memory constraints did not impact benchmark performance. The software environment consisted of Julia version 1.10.9, SymbolicUtils.jl version 3.25.1 and Symbolics.jl version 6.36.0. 

\subsection{Benchmarks}
We varied the size of the mathematical models by increasing the number of variables and equations in each problem. For each model size, we measured the following stages of the symbolic-numeric workflow.
\begin{itemize}
    \item Jacobian Symbolic Computation: Measures time and memory to symbolically compute the Jacobian matrix.
    \item Code Generation: Measures time to generate a compilable representation of the Jacobian function.
    \item Function Compilation: Measures time to compile the generated function using Julia's compiler.
    \item Function Evaluation: Measures the time to evaluate the compiled numerical function.
\end{itemize}

The metrics include time (seconds) and allocated memory (bytes).

\subsection{Results}\label{sec:results}
\begin{figure}
    \centering
    \includegraphics[width=1\linewidth]{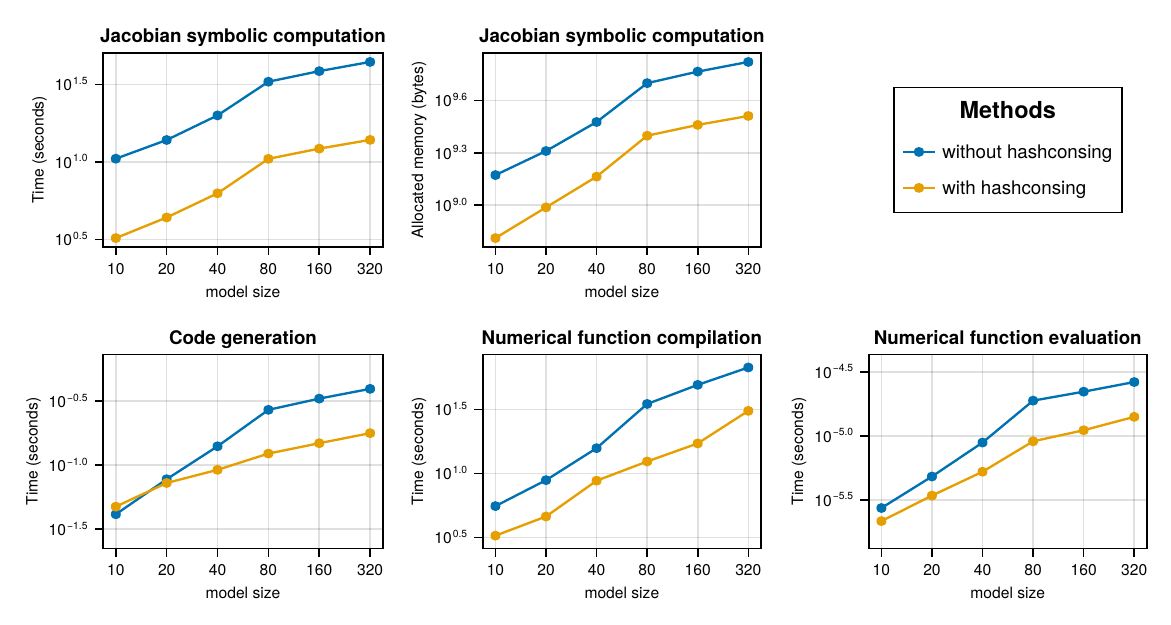}
    \caption{Benchmark results on the B Cell Receptor (BCR) problem (\autoref{sec:BCR}), comparing performance with and without hash consing across symbolic computation, code generation, compilation, and evaluation stages.}
    \label{fig:bcr}
\end{figure}

\begin{figure}
    \centering
    \includegraphics[width=1\linewidth]{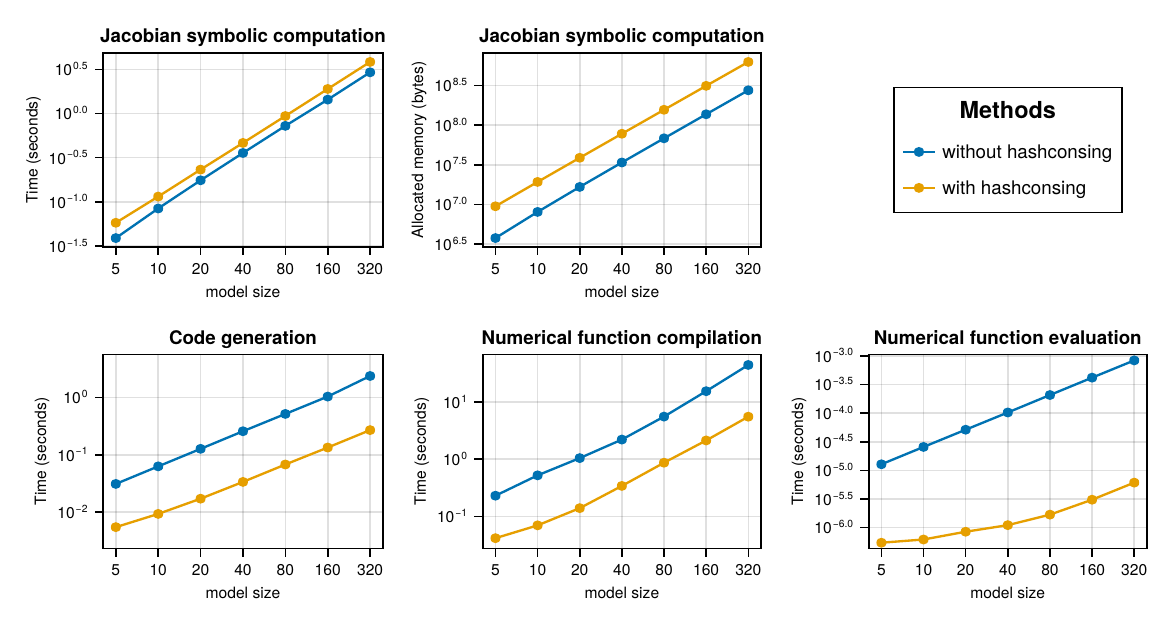}
    \caption{Benchmark results on the X Steam thermodynamic system (\autoref{sec:thermal}), comparing performance with and without hash consing across stages and varying model sizes. Note the significant improvements in compilation and evaluation times.}
    \label{fig:thermal}
\end{figure}

The benchmark results for the BCR model and the X Steam thermodynamic system are presented in \autoref{fig:bcr} and \autoref{fig:thermal}, respectively.

For the BCR model, hash consing demonstrates consistent benefits across most stages. Jacobian symbolic computation time improved by approximately $3\times$, while memory allocation for this stage was reduced by roughly $2\times$. Compilation and numerical evaluation times also saw improvements of around $2\times$. Notably, the code generation step showed nuanced behavior: without hash consing, it was faster for smaller models, but with hash consing enabled, it became faster for larger models. This crossover is attributed to the slight overhead introduced by traversing the Directed Acyclic Graph (DAG) structure created by hash consing, an overhead outweighed by benefits as expression complexity and redundancy grow.

For the XSteam thermodynamic system, the results highlight the importance of considering the entire workflow. During the initial Jacobian symbolic computation, hash consing resulted in slightly slower execution times and increased memory allocation by approximately $2.5\times$. This is likely because the XSteam equations, while complex overall, generate fewer structurally identical sub-expressions involving variables during the differentiation process compared to the BCR model, making the overhead of hash consing (memoization) dominant in this initial phase. However, significant benefits emerged downstream: code generation time improved consistently by around $5\times$, function compilation time improved by nearly 10 times, and numerical function evaluation speedups ranged dramatically from $20\times$ to $100\times$ across different model sizes. This indicates that while initial expression construction might see less benefit or even overhead, the identification of common subexpressions overall (including constants and intermediate calculations) drastically improves the efficiency of subsequent compilation and execution.
\section{Discussion and Generalizability}
The benchmark results presented in \autoref{sec:results} confirm that hash consing can be a highly effective technique for improving the memory efficiency and computational performance of symbolic systems like JuliaSymbolics, though its impact varies depending on the problem structure and the stage of the symbolic-numeric workflow.

\subsection{Analysis of Performance Patterns}
The substantial memory savings observed, particularly in the BCR model's Jacobian computation (reduction by $2\times$, stem directly from the core principle of hash consing: sharing structurally equivalent subexpressions. This mitigates the memory explosion often seen in symbolic manipulation, especially with large, redundant expressions common in ODE systems derived from reaction networks.

Performance improvements show a more complex pattern. The BCR model benefits broadly, with speedups around $2-3\times$ in Jacobian computation, compilation, and evaluation. This is consistent with pointer equality accelerating structural comparisons during term traversal and manipulation inherent in these steps. The code generation crossover highlights that optimizations targeting DAGs have an initial overhead but pay off as scale increases.

Conversely, the XSteam results demonstrate that hash consing's benefits are not always uniform across all stages. The initial overhead during Jacobian computation (slight slowdown, around $2.5\times$ memory increase) underscores that if the initial symbolic manipulation generates few immediately identical variable-based terms, the costs of hashing and lookups can dominate. However, the dramatic downstream speedups in code generation ($5\times$), compilation ($10\times$), and especially evaluation (up to $100\times$) are compelling. This strongly suggests that hash consing effectively identifies and eliminates redundancy in the overall expression structure (including numerical constants and intermediate results), which is then heavily exploited by Julia's compiler during code generation and optimization, leading to much faster numerical execution. This is a significant finding for automated code generation workflows.

These results imply that hash consing is particularly potent in workflows involving large-scale symbolic differentiation, simplification, and subsequent code generation (e.g., Jacobian generation, PDE discretization), common in tools like ModelingToolkit.jl. By significantly reducing memory pressure and accelerating key compilation and runtime phases, hash consing enhances the scalability of symbolic-numeric coupling. The overhead, while measurable in specific scenarios like the initial XSteam Jacobian step, is generally outweighed by substantial gains in GC time reduction and faster term processing, especially for large, complex mathematical models.

Our implementation choices, using \texttt{WeakValueDict} and modifying constructors, proved effective. Weak references successfully managed memory without leaks, though the potential for garbage collection requiring expression recreation represents a minor theoretical overhead not observed to be problematic in practice. Transparent integration via constructors ensured broad applicability within the existing ecosystem.

\subsection{Generalization to Other Systems}
The success of hash consing in JuliaSymbolics suggests potential benefits for other symbolic computation systems, but practical adoption depends on language and ecosystem features. 

\subsubsection{Language and Runtime Characteristics}
The effectiveness of hash consing depends significantly on the host language and runtime environment.

\paragraph{Weak References} 
Languages with support for efficient hash table implementations and weak references are generally better suited for hash consing. Systems in Python or JavaScript can replicate our weak-reference-based table using \texttt{weakref.WeakKeyDictionary} or \texttt{WeakMap}, but global interpreter locks (e.g., Python’s GIL) may limit concurrency.

\paragraph{Garbage Collection Strategy} 
The reliance on weak references means that performance is partly dependent on the garbage collection behavior of the runtime environment. Systems with efficient garbage collectors that minimize fragmentation may be more suitable for hash consing. Languages with tracing garbage collectors (like Julia, Java, Python) benefit more readily from our weak reference approach than those with reference counting (like CPython) or manual memory management. 

\subsubsection{Symbolic System Architecture}
The architectural characteristics of the symbolic system significantly impact the applicability of hash consing.

\paragraph{Expression Immutability} 
The success of hash consing in JuliaSymbolics hinges on Julia’s support for immutable value types and deterministic garbage collection. Mutable symbolic systems (e.g., Mathematica’s stateful expressions \cite{Mathematica}) would require additional mechanisms to maintain hash table integrity.

\paragraph{Constructor Access} 
Hash consing requires intercepting object creation, which is easier in systems with centralized expression constructors than those with distributed creation patterns.

\paragraph{Expression Normalization} 
Systems that normalize expressions to canonical forms (such as sorting commutative operands) achieve higher hash consing hit rates than those without such normalization.

\paragraph{API Design} 
Systems with clear separation between internal representation and external interface can implement hash consing transparently, while those with tightly coupled representations may require API changes.

\paragraph{Algorithm Adaptation for DAGs}
Perhaps most critically, systems must revise their internal algorithms to take full advantage of the DAG structure produced by hash consing. Many symbolic algorithms, such as simplification, substitution, rewriting, and code generation, are traditionally designed to operate on tree-structured expressions. When expressions become maximally shared DAGs, these algorithms must be adapted to recognize shared substructures, avoid redundant traversal, and correctly manage evaluation order. As demonstrated in our modified code generation pass, failing to account for the DAG structure can negate the performance gains of hash consing. Thus, effective adoption of hash consing involves not just memory deduplication but also rethinking the symbolic algorithms themselves.

\subsection{Comparison to existing symbolic libraries}

There are several other symbolic algebra packages, both open- and closed-source, implemented in a variety of languages. SymEngine \cite{symengine} is a commonly used symbolic algebra package written in C++, with a Python wrapper. SymPy \cite{SymPy2017} is a similar package written purely in Python. While SymEngine does not perform any caching, SymPy does cache several expression constructors. Neither of them implement true hash-consing. Constructing identical expressions with different memory locations amounts to comparing a manually expanded polynomial expression with one generated via the \texttt{expand} function in each of the packages. In fact, the common subexpression elimination code in SymPy \cite{SymPyCSESource} and SymEngine \cite{SymEngineCSESource} are implemented by storing identical subexpressions in a set data structure, instead of by leveraging the DAG structure inherent in a hash-consed implementation.

Other solutions such as FriCAS \cite{FriCAS} and REDUCE \cite{REDUCE1974}are implemented in Lisp, and while they benefit from Lisp's interning of symbols and strings, we did not find a proper hash-consing implementation. Notably, the Maxima CAS \cite{maxima} has a thread \cite{maxima2023slow} discussing noticeable slowdowns with hash-consing being proposed as a potential improvement. A thread as recent as 2025 \cite{maxima2025hcons} discusses the potential for hash-consing in Maxima.

The GiNaC \cite{GiNaC} symbolic algebra package does use a form of reference counting for symbolic expressions. Symbolica \cite{symbolica2025} is a new symbolic algebra package implemented in Rust. It is trivial to construct programs using either package that demonstrate identical subexpressions with different memory locations.

Closed-source implementations also leverage hash-consing, though verifying this proves to be difficult. A blog post discussing the internals of the Wolfram language \cite{wolfram_internals} contains descriptions that suggest hash-consing is used internally. However, there is no description of the performance effects this process has and what algorithms leverage it.

\section{Conclusion and Future Work}
\subsection{Conclusion}
This paper has presented the design, implementation, and evaluation of hash consing for efficient symbolic computation in the JuliaSymbolics system. Our work addresses the critical challenge of memory redundancy in symbolic expression representation, which has been a persistent limitation in the scalability and performance of symbolic computation systems.

We have shown that by transforming symbolic expression trees into simplified directed acyclic graphs (DAGs) without duplicate through hash consing, we can achieve significant improvements in both memory efficiency and computational performance. Our implementation in JuliaSymbolics demonstrates:
\begin{itemize}
    \item Memory consumption reductions: Achieved approximately $2\times$ reduction during Jacobian computation for the complex BCR ODE model, while effectively controlling memory growth in scalable thermodynamic models.
    \item Computation time improvements: Demonstrated speedups across various stages, including $3\times$ for Jacobian computation (BCR), up to $5\times$ for code generation (XSteam), nearly $10\times$ for compilation (XSteam), and dramatic speedups up to $100\times$ in numerical function evaluation (XSteam).
    \item Reduced garbage collection pressure: Implied by the significant memory allocation reductions observed, particularly in memory-intensive symbolic construction phases.
    \item Enhanced scalability: Showcased through controlled memory usage and performance gains that become more pronounced with increasing model complexity, especially in the compilation and evaluation of generated code.
\end{itemize}

The integration of hash consing into JuliaSymbolics was accomplished through a transparent approach that maintains API compatibility while fundamentally transforming the underlying expression representation. The use of weak references proved crucial for effective memory management in Julia's garbage-collected environment.

Our findings confirm that hash consing, despite its history, remains a highly relevant and powerful technique for modern, dynamically-typed symbolic systems. The performance profiles observed across different benchmarks highlight that while benefits might vary per stage (with potential minor overhead in initial construction for some problems), the downstream impact on code generation, compilation, and especially numerical execution can be substantial.

This work directly enhances JuliaSymbolics' capabilities in its core domains, differential equations, model transformation, and runtime code generation, by providing a more efficient computational foundation. The memory and performance gains are particularly valuable for the complex, large-scale models prevalent in modern scientific computing and automated analysis workflows. 

\subsection{Future Work}
While our implementation of hash consing provides substantial benefits, several promising directions for future research and development remain.

\subsubsection{Integration with E-Graphs}
A particularly promising direction is the integration of hash consing with equality saturation using e-graphs \cite{e-graph1980}. E-graphs are data structures that represent sets of equivalent expressions, known as e-classes, allowing for simultaneous exploration and rewriting of multiple equivalent forms. By combining hash consing with e-graphs, we can simultaneously leverage the memory efficiency of hash consing and the algebraic simplification power of e-graph rewriting.

In this combined approach, structurally identical expressions would be hash-consed as before, ensuring a unique representation in memory. Furthermore, equivalent expressions within the same e-class, such as \texttt{(a * 2) / 2}, \texttt{(a << 1) / 2}, and \texttt{a} where \texttt{a} is a symbolic symbol that behaves like an integer, would be associated within the e-graph structure. The goal would be to maintain a single, unique copy of the optimal expression within each e-class, based on a cost function that reflects factors such as expression size, computational complexity, or code generation efficiency. In effect, hash consing ensures each subexpression is at a unique location in memory, while an e-graph ensures that each expression is at a unique location in logic (up to equivalence).

E-graphs have been successfully applied in diverse areas, such as compiler optimization \cite{Denali2002,EqualitySaturation2009}. Within symbolic computation, the integration of e-graphs with hash consing is expected to further reduce memory consumption by identifying and sharing logically equivalent expressions in addition to structurally identical ones. Moreover, it can enable more performant code generation by allowing the selection of the most efficient representation for each expression within its e-class. This has the potential to significantly enhance the capabilities of JuliaSymbolics in domains such as scientific computing and model-based engineering, where efficient symbolic manipulation and code generation are paramount. We believe this path is likely to be very fruitful.

\subsubsection{Cross-Thread Caching}
The current implementation employs task-local hash consing caches and memoization tables to ensure lock-free operation in multithreaded contexts. While this design avoids synchronization overhead, when multiple threads compute the same value, redundant work remains to be performed. 

To address this, exploring cross-thread caching presents a promising optimization opportunity. By enabling threads to share cached computation results, redundant calculations can be significantly reduced. For instance, in a parallelized symbolic computation scenario where multiple threads are handling different parts of a large expression, if a sub-expression is common across threads, the result of its computation could be shared.

Implementing cross-thread caching requires careful consideration of several factors. First, the cache management system needs to be thread-safe to prevent data races and ensure the integrity of cached values. This may involve using synchronization mechanisms such as mutexes or more advanced lock-free data structures. Second, the cache invalidation strategy becomes crucial. When an expression is updated or modified in one thread, the corresponding cached values in other threads must be updated or invalidated to maintain consistency.

Moreover, the performance gain from cross-thread caching will depend on the nature of the symbolic computations. In workloads with a high degree of shared sub-expressions across threads, the benefits will be substantial, leading to faster overall computation times and improved resource utilization. However, in cases where thread-specific computations dominate, the overhead associated with cross-thread cache management may outweigh the benefits.

This area of research can be further extended to distributed systems, where multiple processes or nodes could potentially share cached symbolic computation results. This would not only enhance the performance of multi-threaded symbolic computations but also have implications for large-scale scientific computing and AI-driven mathematical reasoning applications that rely on distributed processing.

\subsubsection{Adaptive Hash Consing Strategies}
Not all expressions benefit equally from hash consing. Future work could explore adaptive strategies that apply hash consing selectively based on expression characteristics and dynamically adjust hash consing policies based on runtime metrics.

\subsubsection{Cross-Session Persistence}
For long-running scientific workflows that span multiple sessions, persisting the hash-consed representation between runs could provide additional benefits. In iterative AI-mathematics pipelines, this would allow LLM agents to reuse canonicalized expressions from prior reasoning steps, avoiding redundant term construction across sessions. A shared hash-consed database could also serve as a persistent symbolic knowledge base for LLMs, similar to retrieval-augmented generation (RAG) architectures in NLP.
\newline

These directions build upon the foundation established in this work, offering paths toward more efficient symbolic computation and enhanced synergy with AI-driven reasoning. By addressing the representational challenges of modern workflows, from traditional scientific computing to LLM-based problem-solving, we can expand the applicability of symbolic methods across increasingly complex domains.

\section{Data Availability}
The hash consing functionality discussed in this work is available as part of \href{https://github.com/JuliaSymbolics/SymbolicUtils.jl}{SymbolicUtils.jl} starting from version 3.28.0. The code and results of the two benchmark problems discussed in \autoref{sec:evaluation} are publicly accessible through the SciML Benchmarks repository at
\begin{enumerate}
    \item \url{https://docs.sciml.ai/SciMLBenchmarksOutput/dev/Symbolics/BCR/}
    \item \url{https://docs.sciml.ai/SciMLBenchmarksOutput/dev/Symbolics/ThermalFluid/}
\end{enumerate}

\printbibliography

\end{document}